\documentclass[journal]{IEEEtran}
\usepackage{graphicx}
\usepackage{epsf}
\usepackage{mathtools}
\usepackage{amssymb}
\usepackage{amsmath, amsfonts}
\usepackage{latexsym}
\usepackage{multirow}
\usepackage{multicol}
\usepackage[table]{xcolor}
\usepackage{color}

\usepackage{tabularx}
\usepackage{lipsum}

\usepackage{algorithm}
\usepackage{algorithmic}

\usepackage{mathrsfs}

\usepackage{fixltx2e}

\usepackage[mathscr]{euscript}

\usepackage{commath}
\usepackage{MnSymbol}

\usepackage{subcaption}

\usepackage{mathtools} 
\DeclarePairedDelimiterX\setc[2]{[}{]}{\,#1 \;\delimsize\vert\; #2\,}
\DeclarePairedDelimiterX\parth[2]{(}{)}{\,#1 \;\delimsize\vert\; #2\,}

\PassOptionsToPackage{hyphens}{url}
\usepackage[unicode=true, bookmarks=true, bookmarksnumbered=true, bookmarksopen=true, bookmarksopenlevel=1, breaklinks=false, pdfborder={0 0 0}, pdfborderstyle={}, backref=false, colorlinks=false]{hyperref}

\usepackage{theorem}
{
\theorembodyfont{\rmfamily}
\newtheorem{assumption}{Assumption}
\newtheorem{remark}{Remark}
\newtheorem{lemma}{Lemma}
\newtheorem{proposition}{Proposition}
\newtheorem{definition}{Definition}

}

\definecolor{orange}{RGB}{255,127,0}
\definecolor{blue}{RGB}{0,0,255}
\definecolor{red}{RGB}{255,0,0}
\definecolor{green}{RGB}{50,160,50}
\definecolor{grey}{RGB}{125,120,125}
\definecolor{purple}{RGB}{125,0,125}
\definecolor{brown}{RGB}{150,75,0}

\begin{document}
{
\title{{\fontsize{23}{2}\selectfont On the Byzantine-Fault-Tolerant Consensus for Blockchain among Connected Vehicles}}

\author
{
Seungmo Kim, \textit{Senior Member}, \textit{IEEE}, and Byung-Jun Kim

\vspace{-0.3 in}

\thanks{S. Kim is with the Department of Electrical and Computer Engineering, Georgia Southern University in Statesboro, GA, USA. B. J. Kim is with the Department of Mathematical Sciences, Michigan Technological University in Houghton, MI, USA. The corresponding author is S. Kim who can be reached at seungmokim@georgiasouthern.edu.}
}

\maketitle
\begin{abstract}
It is a critical matter for a blockchain system whether a Byzantine fault tolerance (BFT) can be guaranteed during a consensus process. Can connected vehicles (CVs) achieve the BFT consensus when the vehicles keep mobile? This paper seeks an answer to this fundamental question. It focuses on characterizing the impact of mobility on the performance of a BFT consensus among CVs.
\end{abstract}

\begin{IEEEkeywords}
Blockchain, Consensus, Byzantine fault tolerance, Connected vehicles
\end{IEEEkeywords}

%%%%%%%%%%%%%%%%%%%%%%%%%%%%%%%%%%%%%%%%%%%%%%%%%%%%%%%%%%%%%%%%%%%%%%%%%%%%%%%%%%%%%%%%%%%%%%%%%%%%%%%%%%%%%%%%%%%%
%%%%%%%%%%%%%%%%%%%%%%%%%%%%%%%%%%%%%%%%%%%%%%%%%%%%%%%%%%%%%%%%%%%%%%%%%%%%%%%%%%%%%%%%%%%%%%%%%%%%%%%%%%%%%%%%%%%%
\section{Introduction}\label{sec_intro}
\subsubsection{Background}
Throughout the short yet rich history of connected vehicles (CVs), application of blockchain has recently taken the center stage in the evolution. A salient paradigm in this realm is how to reach a consensus efficiently with the presence of mobility and dynamicity that a network of CVs inevitably draws.

Every blockchain system employs a consensus mechanism through which the participating peers reach an agreement on verification of a block \cite{nakamoto08}. In such a distributed consensus, the \textit{Byzantine fault} arises as a challenge, which refers to a condition where some of the participants may fail and hence the consensus should be made among the other participants.

While there has been impetus of prior work on the \textit{Byzantine fault tolerance (BFT)}, this paper puts a particular focus on the feasibility of achieving a BFT consensus in a blockchain system formed among CVs. As mentioned earlier, this environment exhibits particular challenges due to the mobility of vehicles: the composition of nodes keeps varying \cite{access19}.

\subsubsection{Related Work}
One of this paper's authors has done fundament studies that can be regarded as the basis of this work \cite{access19}\cite{fabric20} where the impact of mobility on the performance of block exchange in a vehicular network \cite{access19} and the feasibility of permissioned blockchain established on the Hyperledger Fabric \cite{fabric20} were studied.

Yet small, there exists a body of prior work focusing on the BFT consensus in a vehicular network. A particular consensus mechanism called the ``proof of eligibility'' was proposed as a means to promote the BFT in consensus among CVs \cite{acm19}. Another recent study introduced the federated learning for achieving a BFT consensus among autonomous vehicles \cite{tvt21}.

While having addressed the timely issue, the related work still leaving a key question unresolved: \textit{How the mobility affects the performance of BFT in reaching a consensus?} As such, the literature is still remaining premature in understanding the fundamental performance of accomplishing a BFT consensus across a network of CVs that dynamically change the network characteristics.

\subsubsection{Contributions}\label{sec_intro_contributions}
In the view of the limitation, this paper is dedicated to \textit{precisely measuring} the feasibility of a BFT consensus in a blockchain network formed among CVs \textit{while they are mobile}. Specifically, it lays out a stochastic analysis framework that is dedicated to characterizing:
\begin{itemize}
\item The number of nodes required to form a BFT consensus
\item The latency of a BFT consensus
\end{itemize}
which shall be formally established as Lemmas \ref{lemma_N} and \ref{lemma_latency}, respectively, in Section \ref{sec_analysis}.

\begin{figure}
\vspace{-0.1 in}
\centering
\includegraphics[width = 0.7\linewidth]{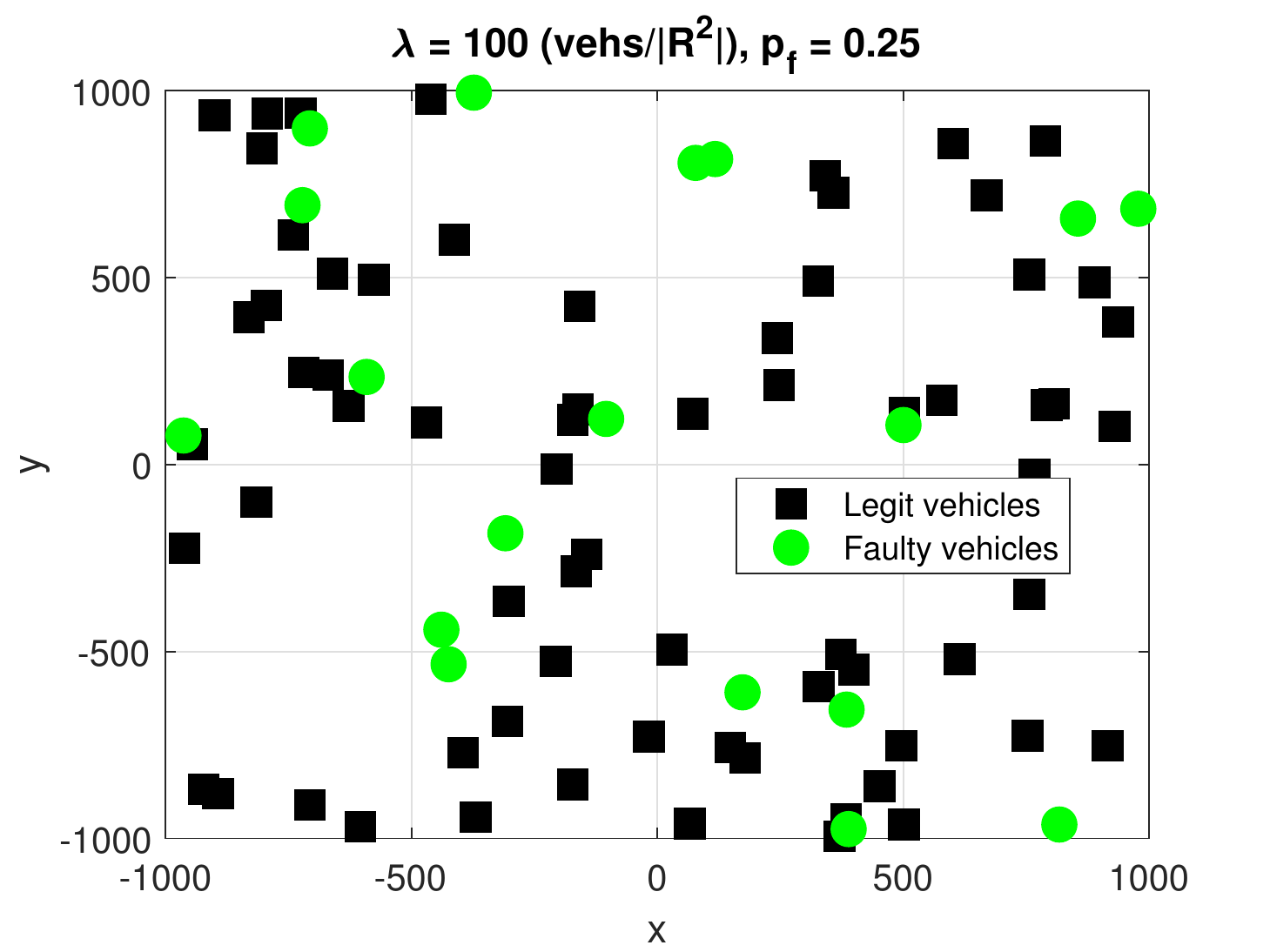}
\caption{An example snapshot of vehicles dropped in $\mathbb{R}^2$ ($\lambda_{N} = 100$ vehicles/$\left|\mathbb{R}^2\right|$ and $p_{f} = 0.25$)}
\label{fig_veh_distribution}
\vspace{-0.2 in}
\end{figure}

%%%%%%%%%%%%%%%%%%%%%%%%%%%%%%%%%%%%%%%%%%%%%%%%%%%%%%%%%%%%%%%%%%%%%%%%%%%%%%%%%%%%%%%%%%%%%%%%%%%%%%%%%%%%%%%%%%%%
%%%%%%%%%%%%%%%%%%%%%%%%%%%%%%%%%%%%%%%%%%%%%%%%%%%%%%%%%%%%%%%%%%%%%%%%%%%%%%%%%%%%%%%%%%%%%%%%%%%%%%%%%%%%%%%%%%%%
\section{System Model}\label{sec_model}
A generalized ``\textit{square}'' space is assumed instead of a road-like setting such as a road segment, for the most generic form of analysis as done in related literature such as \cite{arxiv2005}. The environment represented by the system
space $\mathbb{R}^{2}$, which is defined on a rectangular coordinate with the width and length of $D$ m. Therein, a CV network is defined as a homogeneous Poisson point process (PPP), denoted by $\Phi$, with the density of $\lambda (> 0)$. The position of vehicle $i$ is denoted by $\mathbf{x}_{i} = (x_{i}, y_{i}) \in \mathbb{R}^{2}$. Note also that the PPP discussed in this paper is a stationary point process where the density $\lambda$ remains constant according to different
points in $\mathbb{R}^{2}$.

The horizontal and vertical positions of a vehicle are uniform random variables in $\mathbb{R}^2$, which is the mathematical rationale leading the vehicle distribution to a PPP \cite{haengi05}. Fig. \ref{fig_veh_distribution} illustrates an example drop of vehicles following a PPP with the intensity of $\lambda = 100$. Black squares indicate ``\textit{legitimate}'' vehicles, while green circles give ``\textit{faulty} vehicles. The meanings of the two vehicle types will be elaborated in Section \ref{sec_analysis_bft}.

Not all $N$ nodes are in each other's physical communication range, which necessitates the \textit{propagation} via multiple hops for a consensus when a new block is generated, as shall be elaborated in Section \ref{sec_analysis_latency}.

The CVs form a \textit{permissionless blockchain} system where \textit{all} the nodes participants in a process of verifying a newly generated block. There is no master/slave type of hierarchy among the nodes nor any selection.

%%%%%%%%%%%%%%%%%%%%%%%%%%%%%%%%%%%%%%%%%%%%%%%%%%%%%%%%%%%%%%%%%%%%%%%%%%%%%%%%%%%%%%%%%%%%%%%%%%%%%%%%%%%%%%%%%%%%
%%%%%%%%%%%%%%%%%%%%%%%%%%%%%%%%%%%%%%%%%%%%%%%%%%%%%%%%%%%%%%%%%%%%%%%%%%%%%%%%%%%%%%%%%%%%%%%%%%%%%%%%%%%%%%%%%%%%
\section{Stochastic Analysis on BFT Consensus}\label{sec_analysis}
Each of the following subsections provides technical details on the two contributions identified in Section \ref{sec_intro_contributions}.

%%%%%%%%%%%%%%%%%%%%%%%%%%%%%%%%%%%
\subsection{Number of Nodes Required for BFT Consensus}\label{sec_analysis_bft}
It is well acknowledged that the minimum number of nodes for a BFT consensus is $N = 3f+1$ where $f$ gives the number of nodes that either (i) fails or (ii) maliciously acts \cite{fire08}. Now, the question is \textit{how this number has to change in a network of CVs}. The point of the question is that the vehicles constituting the network are mobile, and hence that $N$ and $f$ have to be quantified in such a way to reflect the mobility.

\begin{proposition}
(Distribution of $f$). \textit{Given the nodes are distributed following a PPP, the number of nodes $N$ in the system area $\mathbb{R}^2$ follows a Poisson distribution with the intensity of $\lambda$. Provided this, the number of faulty nodes $f$ follows another Poisson distribution with the intensity of $p_{f}\lambda$ where $p_{f}$ gives the probability of a failure for each node.}
\end{proposition}

\textit{Proof:}
The proposition can easily be proved predicated on \textit{thinning} of a PPP \cite{book}.
\hfill$\blacksquare$

\vspace{0.2 in}
One of the key contributions of this paper is to introduce new variables characterizing the uniqueness of a network composed of connected vehicles. We focus on the fact that the number of nodes constituting a network \textit{varies} due to the mobility. We introduce two random variables $\delta_{N}$ and $\delta_{f}$ indicating the change in the number of legitimate and faulty nodes, respectively.

Fig. \ref{fig_variables} illustrates the relationship among the variables that are used in the formulation of this paper. We start with reiterating that $N$ and $f$ are Poisson random variables, as has been discussed in Section \ref{sec_model}. Now, as being a CV network, the key characteristic is that the intensities of both $N$ and $f$ vary due to the mobility of vehicles composing the network. We denote the in-/decrement of $N$ by $\delta_{N} = N^{+} - N^{-}$. The same formulation applies to $f$, which yields $\delta_{f} = f^{+} - f^{-}$.

\begin{assumption}\label{assumption_mm1}
(Arrival and departure of vehicles). \textit{Assume an M/M/1 queue. It yields that both arrival and departure processes follow a Poisson distribution based on the Burke's theorem \cite{burke}.}
\end{assumption}

\begin{lemma}\label{lemma_N}
(Distribution of $N$). \textit{The variable $N$ denoting the total number of vehicles required to accomplish a BFT consensus follows another Poisson distribution with the intensity of $3\lambda_{f} - \lambda_{\delta_{N}} + \lambda_{\delta_{f}}$.}
\end{lemma}

\textit{Proof:} Considering the mobility of the vehicles, the number of vehicles to keep the BFT among connected vehicles is computed as
\begin{align}\label{lemma_N_proof}
N > 3f - \delta_{N} + \delta_{f} + 1
\end{align}
Notice that we define $\delta_{N}$ and $\delta_{f}$ as random variables indicating the ``cumulative'' numbers of legit/faulty vehicles flowed into/out of the network over the last $T$ seconds.

Since the Poisson distribution is \textit{closed under addition}, the $N$ given in (\ref{lemma_N_proof}) also follows a Poisson distribution whose intensity is $3\lambda_{f} - \lambda_{\delta_{N}} + \lambda_{\delta_{f}}$.
\hfill$\blacksquare$

\vspace{0.1 in}
\begin{remark}
(Meaning of $N$). \textit{$N$ is the ``required'' number of nodes for existence of $f$ faulty nodes, which translates to ``when there are $f$ faulty nodes, how many nodes does one need to keep the network BFT?'' As such, a network is composed of $N$ nodes where only $\left(N-f\right)/N$ nodes are capable of participating in a consensus process.}
\end{remark}

\begin{figure}
\vspace{-0.1 in}
\centering
\includegraphics[width = 0.6\linewidth]{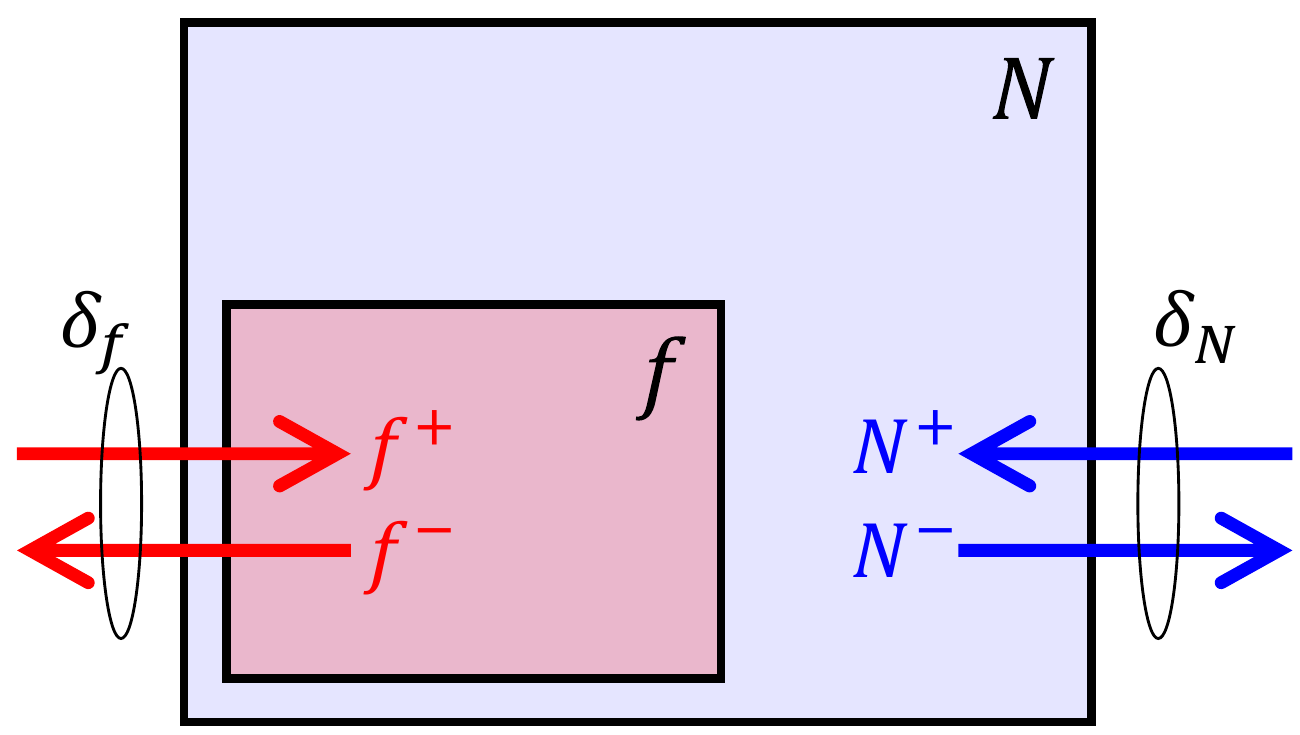}
\caption{Inter-relationship among the variables constituting Lemma \ref{lemma_N}}
\label{fig_variables}
\vspace{-0.1 in}
\end{figure}

%%%%%%%%%%%%%%%%%%%%%%%%%%%%%%%%%%%
\subsection{Latency of BFT Consensus}\label{sec_analysis_latency}
Many of recent blockchain systems adopt the \textit{gossip protocol} in their consensus mechanism \cite{gossip_access19}\cite{gossip_arxiv18}. In what follows, we refer to the epidemiology \cite{gossip11} as a method to formulate how a block is disseminated throughout a network via a gossip protocol.

Here are key variables. We remind that $N$ denotes the total number of CVs in the network. Now, by $r_{t}$ and $\overline{r}_{t}$, we denote the proportion of vehicles that have ``\textit{received}'' and ``\textit{not received}'' the block sent from the source node in the $t$th time slot, which yields $\overline{r}_{t} = 1 - r_{t}$. We assume that in the first time slot, i.e., $t=0$, the source node has a block to propagate, which is given by $\overline{r}_{0} = p_{f}$ where $p_{f}$ gives the probability that a given node fails in passing a block along. As a result of the definition, $r_{0} = 1 - p_{f}$ is true.

\begin{proposition}\label{proposition_rate}
(Block dissemination rate). \textit{With $p_{f}$ denoting the probability that an arbitrary node fails to receive a block, we define the probability that any given node has successfully received a block for a consensus in time slot $t$ as}
\begin{align}\label{eq_proposition_rate}
\mathsf{R} = 1 - \overline{r}_{t-1} p_{f}^{N\left(1 - p_{f}\right)}
\end{align}
\end{proposition}

\textit{Proof:}
We start with formulating the probability that a node fails to receive a block in time slot $t+1$, which can be writen as
\begin{align}\label{eq_gossip}
\overline{r}_{t} &= \mathbb{P}\left[\text{No reception as of } t-1\right]\mathbb{P}\left[\text{No further reception}\right]\nonumber\\
&= \overline{r}_{t-1} p_{f}^{N\left(1 - p_{f}\right)}
\end{align}
where $p_{f}$ denotes the probability that a node fails. We also clarify that $\mathbb{P}\left[\text{No reception as of } t-1\right] = \overline{r}_{t-1}$ and $\mathbb{P}\left[\text{No further reception}\right] = p_{f}^{N\left(1 - p_{f}\right)}$. As such, the equation (\ref{eq_gossip}) translates to the following: in order for a node to remain having no block received at time $t$, it must have received no block at time $t-1$, either (i.e., $\overline{r}_{t-1}$); then, at time $t$, there has to be another failure of block delivery (i.e., $p_{f}$) from all the other nodes that are capable of sending a block (i.e., $N\left(1 - p_{f}\right)$).

This can easily be translated to the probability that a node receives a block in time slot $t$, which is given by
\begin{align}
\mathsf{R} := r_{t} = 1 - \overline{r}_{t},
\end{align}
which completes the proof.
\hfill$\blacksquare$

%\vspace{0.2 in}
\begin{definition}\label{definition_latency}
(Latency of block dissemination). \textit{We define the latency of dissemination of a block throughout a network of CVs as the number of time slots consumed through the course of a gossip protocol until the rate of block dissemination reaches a 1, which is formulated as}
\begin{align}
\mathsf{T} = t|_{\left|r_{t} - 1\right| = \epsilon}
\end{align}
\textit{where $\epsilon$ gives an infinitesmall number that is smaller than the size of a granule used in the computer. We choose $\epsilon = 10^{-5}$ for the results that will be presented in Section \ref{sec_results}.}
\end{definition}

\vspace{-0.1 in}
\begin{lemma}\label{lemma_latency}
(Distribution of latency of BFT consensus among CVs). \textit{The distribution of latency is approximated to the beta distribution.}
\end{lemma}

\textit{Proof:} The latency predominantly depends on the in-/outflow of legit/faulty vehicles, i.e., $\delta_{N}$ and $\delta_{f}$. A beta random variable uniquely models a situation where a pair of \textit{binary} parameters determine the shape of its PDF.

While this proof does not provide a closed-form expression, the suitability of beta distribution is presented in Figs. \ref{fig_latency_small_f} and \ref{fig_latency_large_f} in Section \ref{sec_results}.
\hfill$\blacksquare$

\begin{figure}
\centering
\includegraphics[width = 0.75\linewidth]{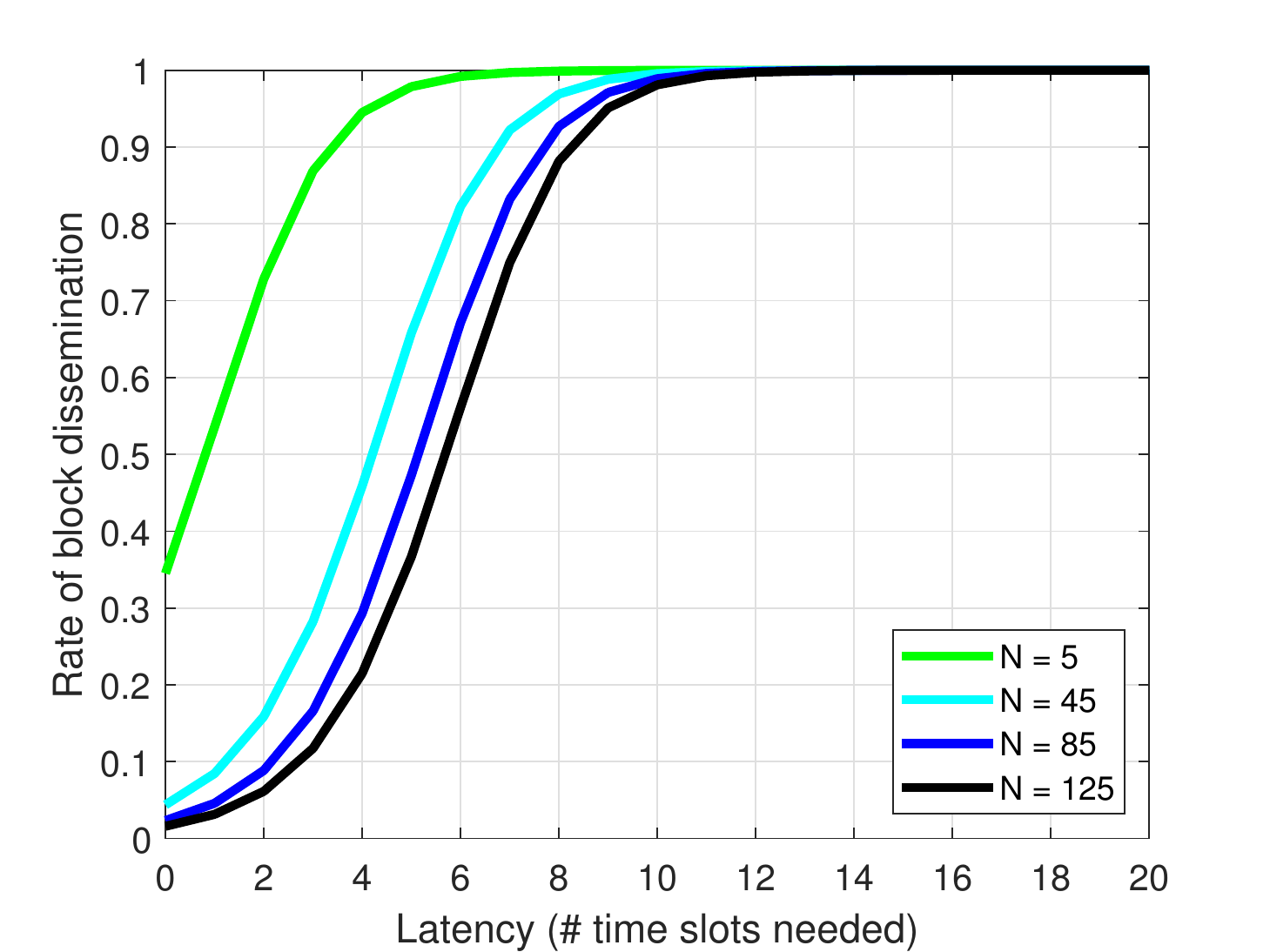}
\caption{Rate of block dissemination via a gossip protocol according to the total number of vehicles $N = \{5,45,85,125\}$}
\label{fig_gossip}
\end{figure}

\begin{figure}
\centering
\begin{subfigure}{0.75\linewidth}
\centering
\includegraphics[width=\linewidth]{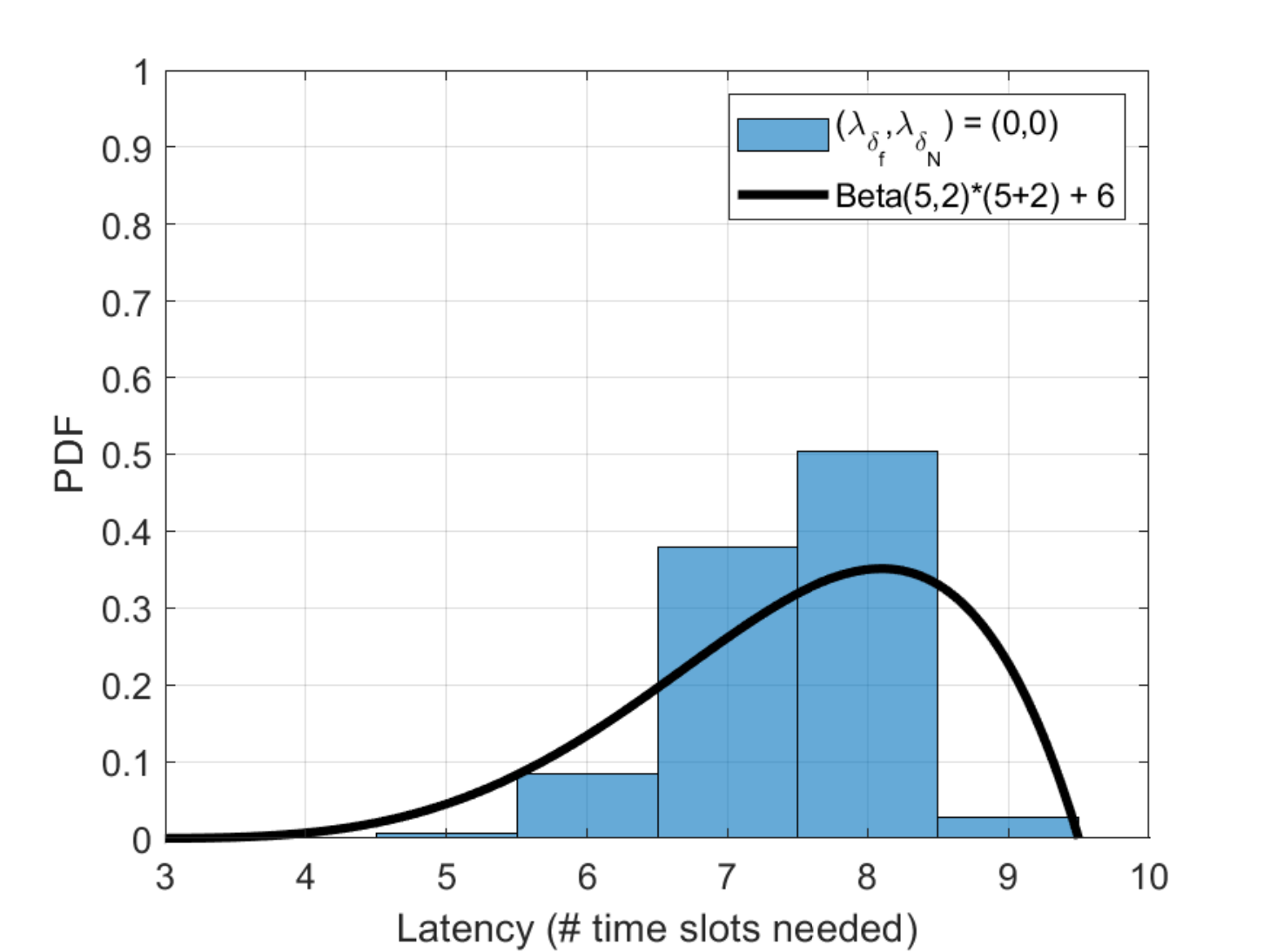}
\caption{Base $f$: $\delta_{f} = \delta_{N} = 0$}
\label{fig_latency_small_f_f}
\end{subfigure}
\hfill
\begin{subfigure}{0.75\linewidth}
\centering
\includegraphics[width=\linewidth]{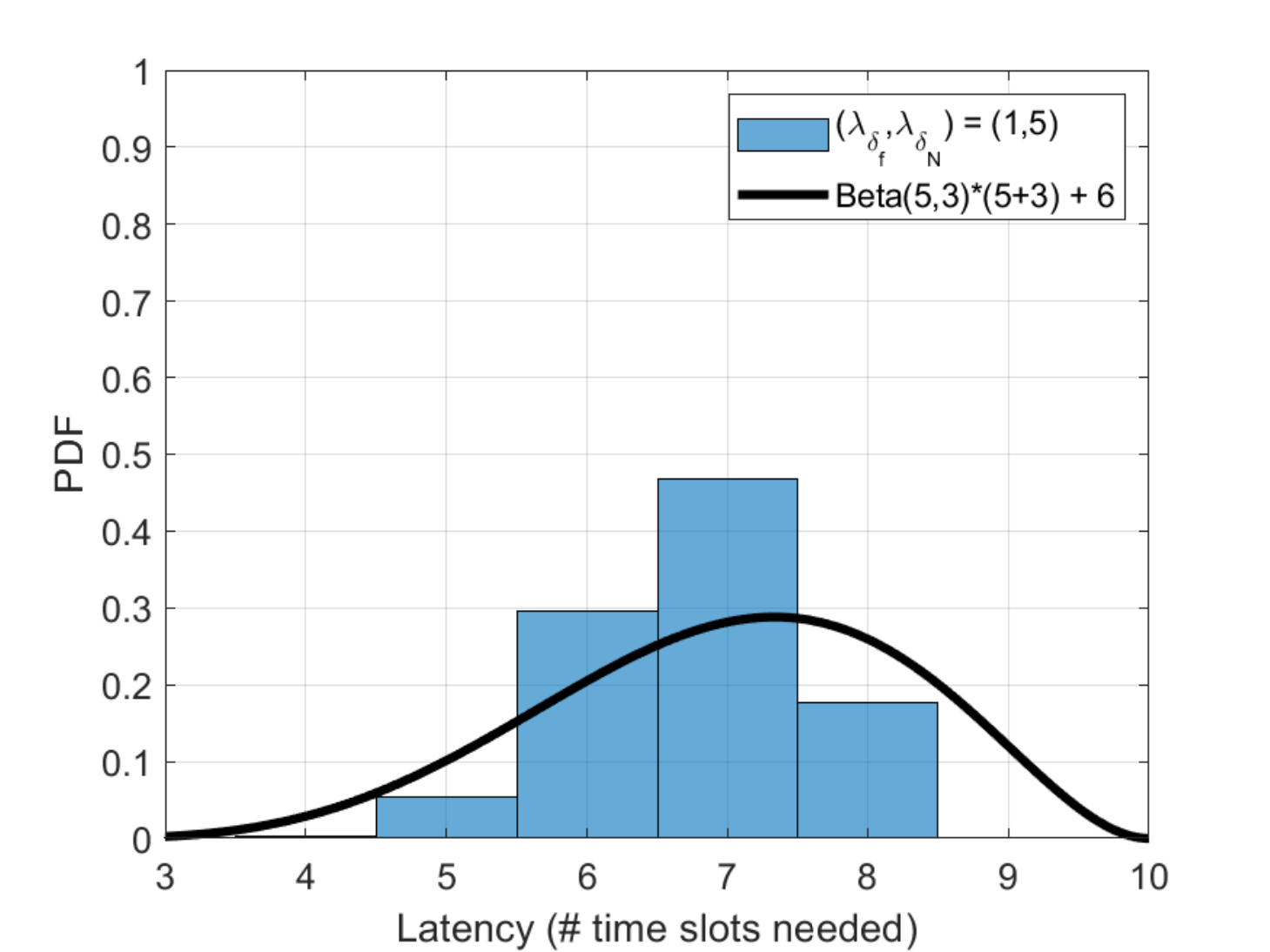}
\caption{$\delta_{f} < \delta_{N}$}
\label{fig_latency_small_f_delta_N}
\end{subfigure}
\hfill
\begin{subfigure}{0.75\linewidth}
\centering
\includegraphics[width=\linewidth]{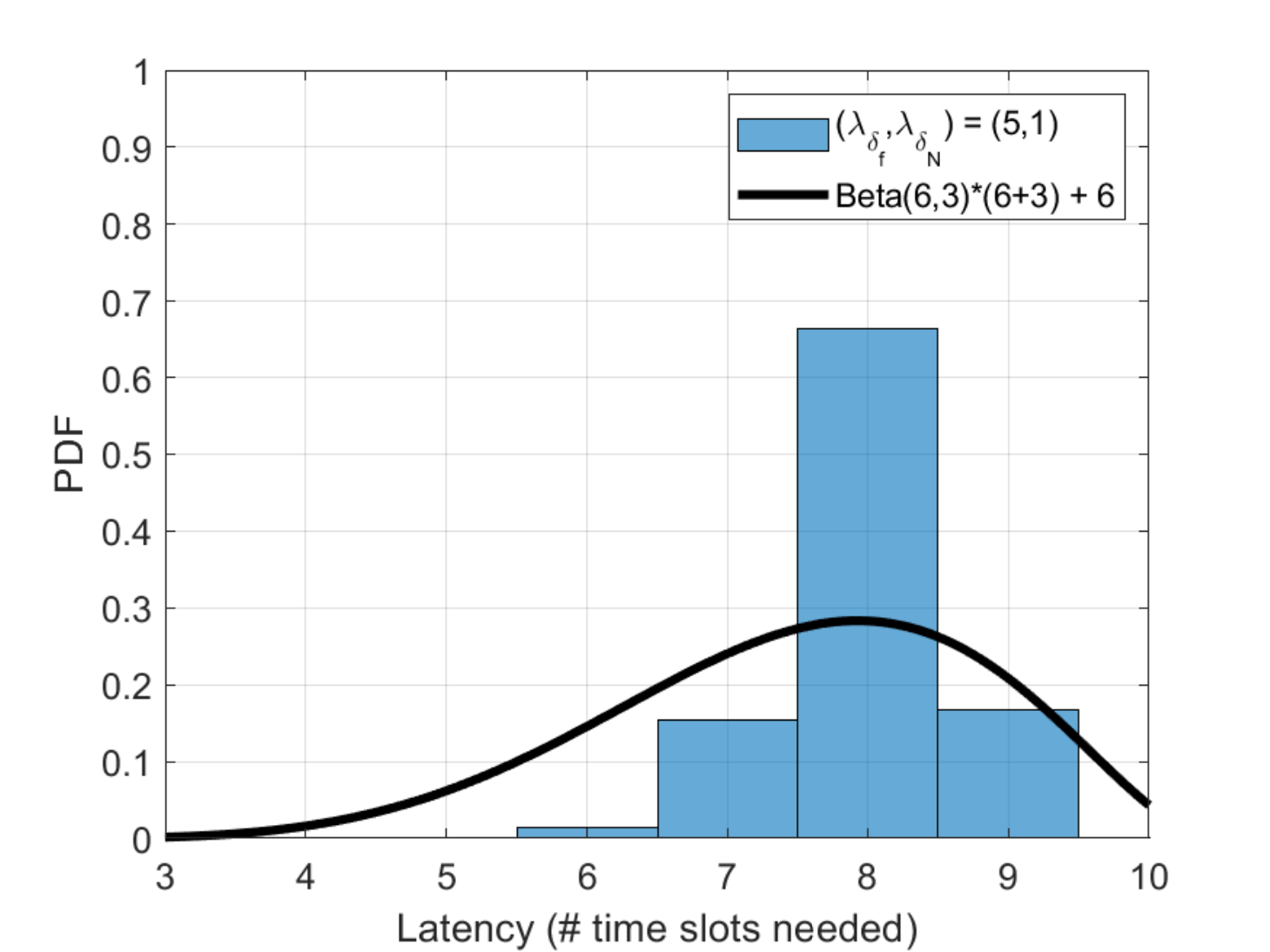}
\caption{$\delta_{f} > \delta_{N}$}
\label{fig_latency_small_f_delta_f}
\end{subfigure}
\caption{Distribution of latency with a small $f$ (i.e., $f=6$)}
\label{fig_latency_small_f}
\end{figure}

\begin{figure}
\centering
\begin{subfigure}{0.75\linewidth}
\centering
\includegraphics[width=\linewidth]{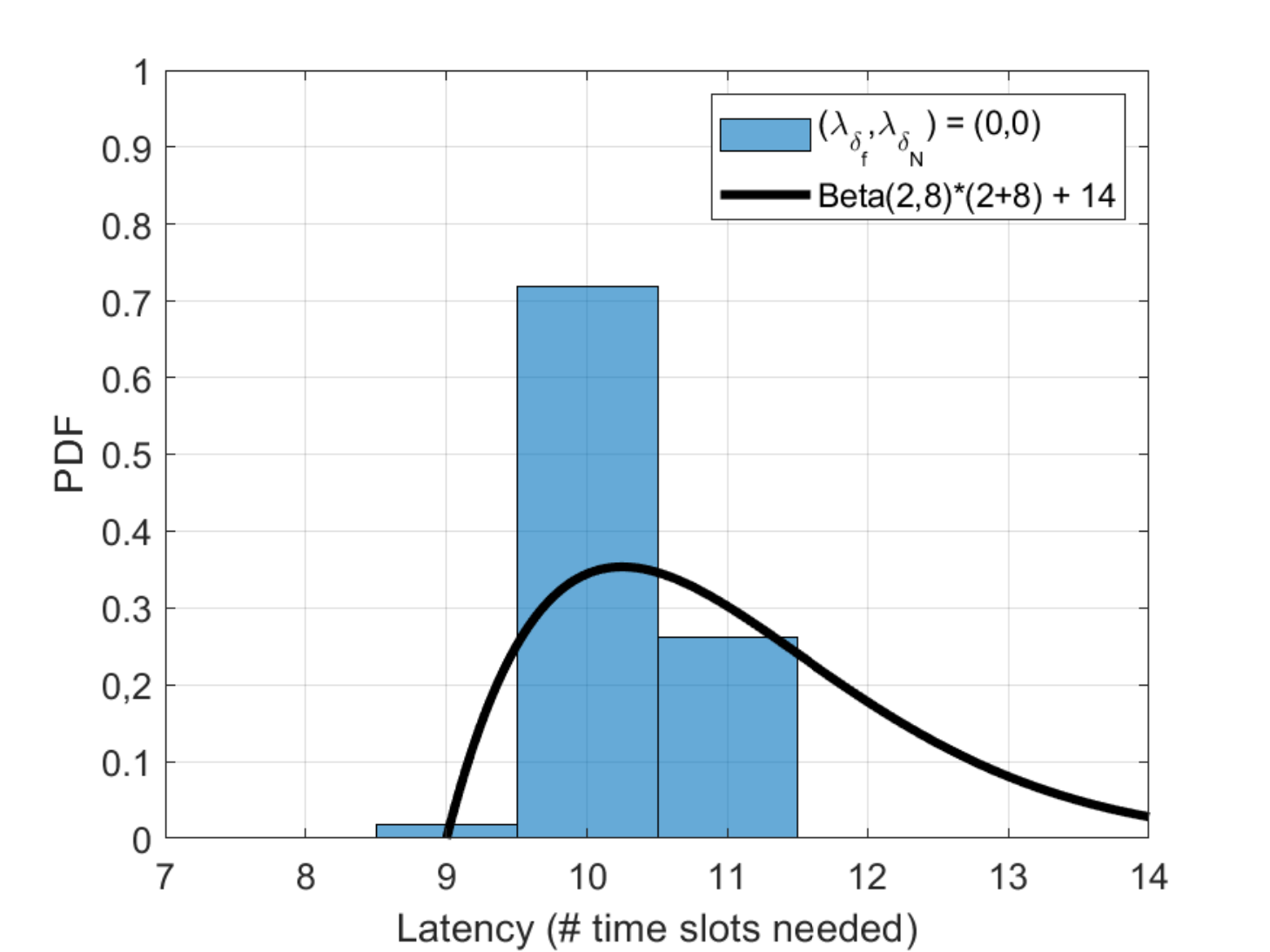}
\caption{Base $f$: $\delta_{f} = \delta_{N} = 0$}
\label{fig_latency_large_f_f}
\end{subfigure}
\hfill
\begin{subfigure}{0.75\linewidth}
\centering
\includegraphics[width=\linewidth]{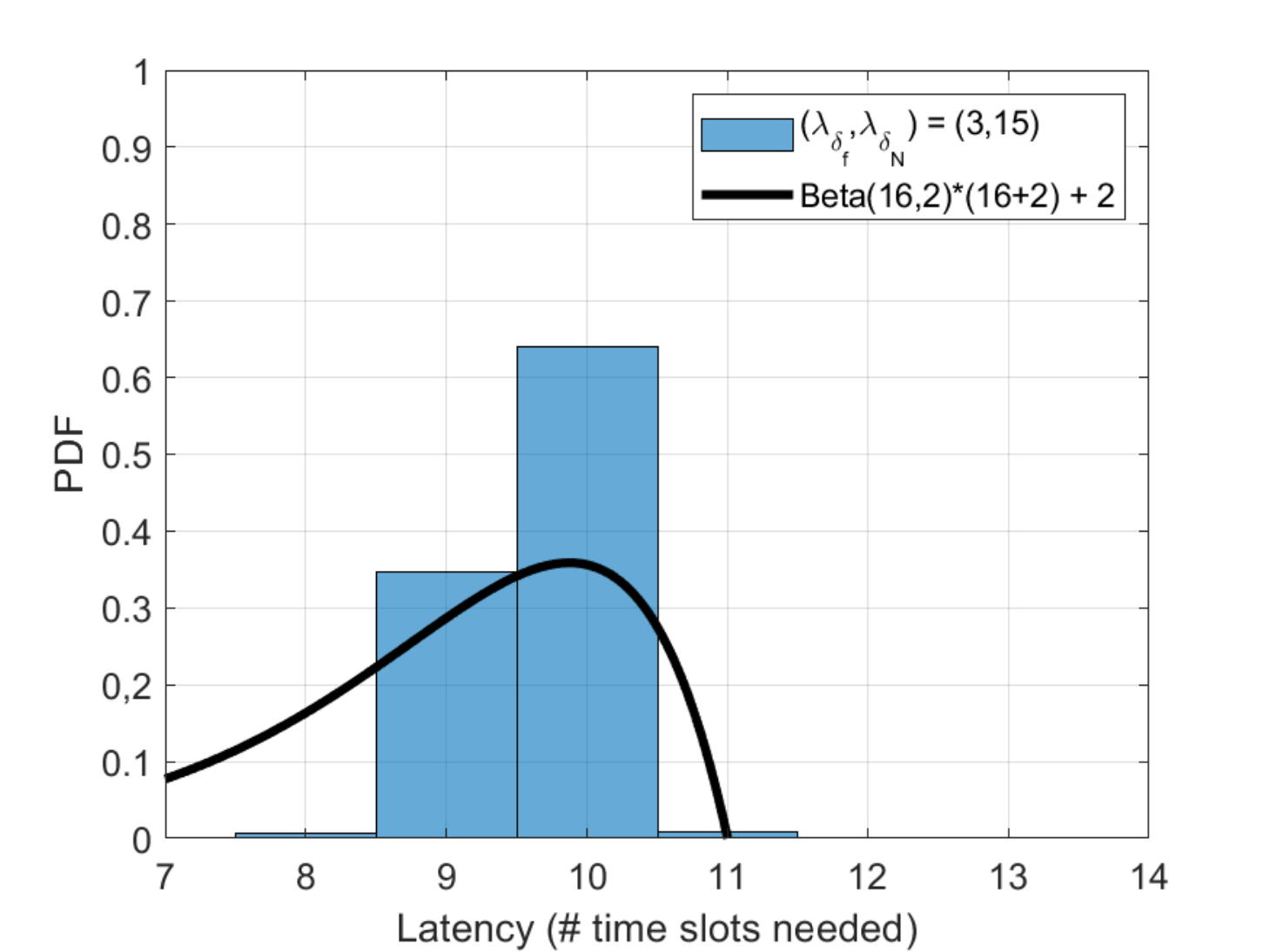}
\caption{$\delta_{f} < \delta_{N}$}
\label{fig_latency_large_f_delta_N}
\end{subfigure}
\hfill
\begin{subfigure}{0.75\linewidth}
\centering
\includegraphics[width=\linewidth]{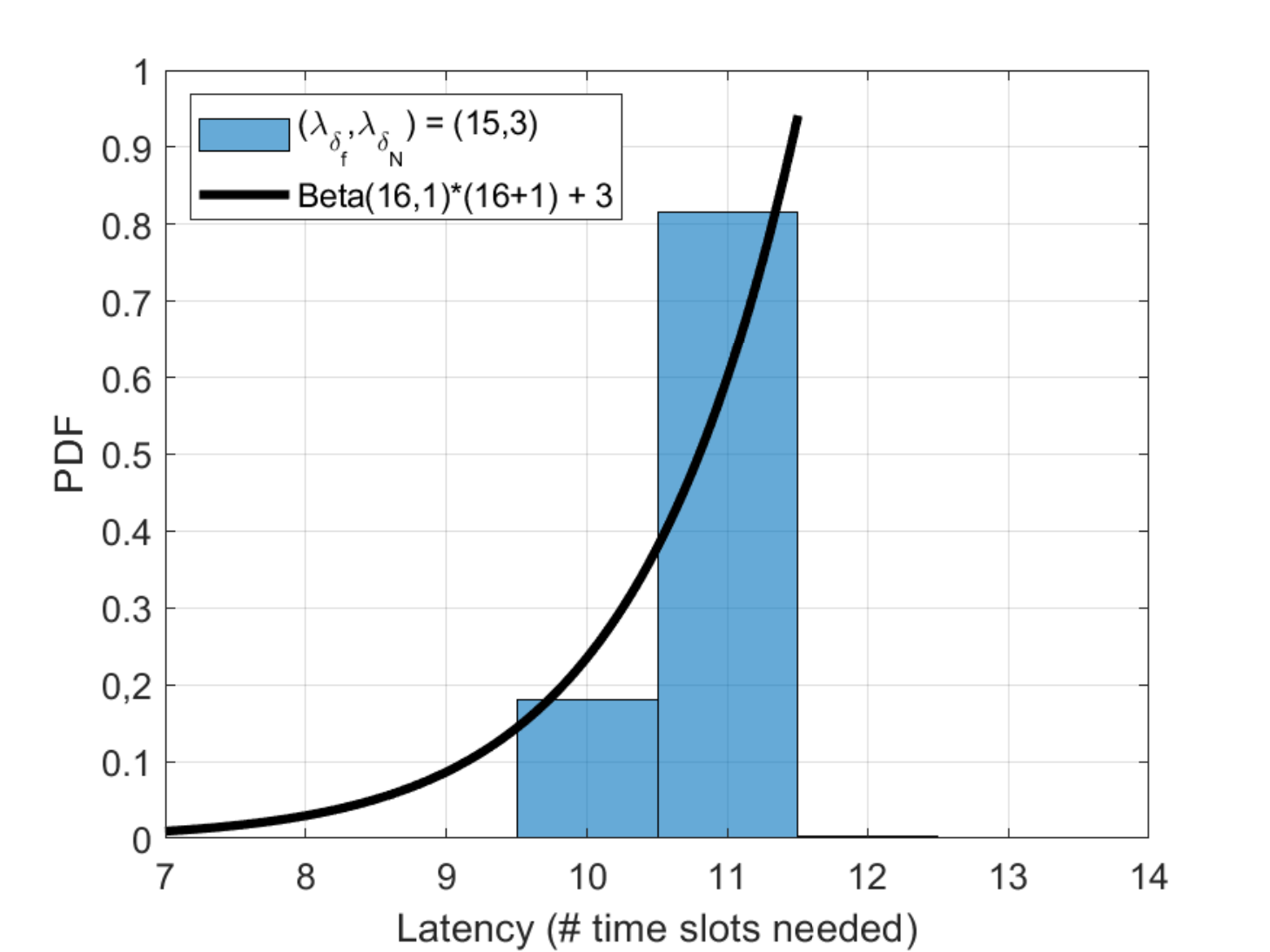}
\caption{$\delta_{f} > \delta_{N}$}
\label{fig_latency_large_f_delta_f}
\end{subfigure}
\caption{Distribution of latency with a large $f$ (i.e., $f=18$)}
\label{fig_latency_large_f}
\end{figure}

%%%%%%%%%%%%%%%%%%%%%%%%%%%%%%%%%%%%%%%%%%%%%%%%%%%%%%%%%%%%%%%%%%%%%%%%%%%%%%%%%%%%%%%%%%%%%%%%%%%%%%%%%%%%%%%%%%%%
%%%%%%%%%%%%%%%%%%%%%%%%%%%%%%%%%%%%%%%%%%%%%%%%%%%%%%%%%%%%%%%%%%%%%%%%%%%%%%%%%%%%%%%%%%%%%%%%%%%%%%%%%%%%%%%%%%%%
\section{Results and Discussion}\label{sec_results}

\subsection{Rate of Block Dissemination}\label{sec_results_rate}
Fig. \ref{fig_gossip} shows the rate of dissemination of a block within a network of CVs, which was formulated as Proposition \ref{proposition_rate}. Each curve represents a value of $N$ on the length of time taken for propagation of a block from the block-generating vehicle to the other vehicles. For instance, the green curve (which is on the very top among the 4 curves) hits 1 when the number of slots consumed is 8. It means that $N = 5$ CVs need 8 time slots for completion of a block propagation among all of the 5 vehicles. The dissemination latency ranges from 8 to 12 slots depending on the given numbers of peers--i.e., $N = \{5, 45, 85, 125\}$.

Furthermore, it is critical to recall that a consensus consumes the largest proportion in the latency for a block from being generated to being finally added to the chain \cite{scale18}. That is, it is useful to understand the latency of a BFT consensus predominates as an effort to fathom the impact of $N$ on the latency of an entire process of verifying a new block.

\subsection{Distribution of Latency of BFT Consensus}\label{sec_results_latency}
Figs. \ref{fig_latency_small_f} and \ref{fig_latency_large_f} describe the distribution of latency $\mathsf{T}$, which was formulated as Lemma \ref{lemma_latency}. The two figures provide a comparative look at how the distribution is differentiated according to the number of faulty nodes, $f$. Further, each figure is composed of three different scenarios: namely, (a) no change in the numbers of legit and faulty nodes: i.e., $\delta_{f} = \delta_{N} = 0$; (b) addition of more legit nodes than faulty nodes: i.e., $\delta_{f} < \delta_{N}$; and (c) addition of more faulty nodes than legit nodes: i.e., $\delta_{f} > \delta_{N}$.

It is also significant to note that each of the figures provides a comparison between (i) the distribution of latency $\mathsf{T}$ and (ii) plot of a best-fitting PDF for a beta distribution, which are shown as a histogram and a black curve, respectively.

Comparison between Figs. \ref{fig_latency_small_f} and \ref{fig_latency_large_f} clearly manifests that a larger $f$ causes a higher latency until a consensus is reached.

Looking into each of Figs. \ref{fig_latency_small_f} and \ref{fig_latency_large_f}, a scenario of $\delta_{f} < \delta_{N}$ generally leads to a lower latency, while that of $\delta_{f} > \delta_{N}$ causes a higher latency.

As such, two key factors elevating the latency of consensus have been found: (i) a larger $f$ and (ii) $\delta_{f} > \delta_{N}$.

It is also noteworthy that in Figs. \ref{fig_gossip} through \ref{fig_latency_large_f} the latency is given in the number of time slots consumed for propagation of a block. We stress that one can easily interpret this latency into the unit of seconds. For instance, in the distributed mode of a C-V2X network (i.e., mode 4), a message is broadcast every \{50, 100, 200\} milliseconds (msec) \cite{vtc21}. This translates a ``\textit{time slot}'' shown in Figs. \ref{fig_gossip} through \ref{fig_latency_large_f} to \{50, 100, 200\} msec. For instance, $\lambda_{N} = 6$ yields that the majority of consensuses are formed with a latency of \{250,500,1000\} msec. The same can be applied to the dedicated short-range communications (DSRC), which sets the inter-message time to 100 msec \cite{access20}.

%%%%%%%%%%%%%%%%%%%%%%%%%%%%%%%%%%%%%%%%%%%%%%%%%%%%%%%%%%%%%%%%%%%%%%%%%%%%%%%%%%%%%%%%%%%%%%%%%%%%%%%%%%%%%%%%%%%%
%%%%%%%%%%%%%%%%%%%%%%%%%%%%%%%%%%%%%%%%%%%%%%%%%%%%%%%%%%%%%%%%%%%%%%%%%%%%%%%%%%%%%%%%%%%%%%%%%%%%%%%%%%%%%%%%%%%%
\section{Conclusions}\label{sec_conclusions}
This paper studied the problem of achieving a BFT consensus among CVs constituting a blockchain network. It provided a stochastic analysis on (i) the number of vehicles to guarantee a BFT consensus and (ii) the latency in forming a consensus. Through simulation, we confirmed the analytic findings and discovered that both the (i) number and (ii) inflowing number of faulty nodes predominate the latency.

%%%%%%%%%%%%%%%%%%%%%%%%%%%%%%%%%%%%%%%%%%%%%%%%%%%%%%%%%%%%%%%%%%%%%%%%%%%%%%%%%%%%%%%%%%%%%%%%%%%%%%%%%%%%%%%%%%%%
%%%%%%%%%%%%%%%%%%%%%%%%%%%%%%%%%%%%%%%%%%%%%%%%%%%%%%%%%%%%%%%%%%%%%%%%%%%%%%%%%%%%%%%%%%%%%%%%%%%%%%%%%%%%%%%%%%%%


\begin{thebibliography}{99}
\setlength{\parskip}{0.0000001 em}

% Introduction
\bibitem{nakamoto08} S. Nakamoto, ``Bitcoin: A peer-to-peer electronic cash system,'' Oct. 2008. [Online]. Available: \url{https://bitcoin.org/bitcoin.pdf}

\bibitem{access19} S. Kim, ``Impacts of mobility on performance of blockchain in VANET,'' \textit{IEEE Access}, vol. 7, May 2019.

\bibitem{fabric20} S. Kim and A. S. Ibrahim, ``Byzantine-fault-tolerant consensus via reinforcement learning for permissioned blockchain implemented in a V2X network,'' \textit{arXiv:2007.13957}, Jul. 2020.

% Related Work
\bibitem{acm19} H. Liu, C. W. Lin, E. Kang, S. Shiraishi, and D. M. Blough, ``A Byzantine-tolerant distributed consensus algorithm for connected vehicles using proof-of-eligibility,'' in \textit{Proc. ACM MSWiM 2019}.

\bibitem{tvt21} J. H. Chen, M. R. Chen, G. Q. Zeng, and J. S. Weng, ``BDFL: a byzantine-fault-tolerance decentralized federated learning method for autonomous vehicle,'' \textit{IEEE Trans. Veh. Technol.} vol. 70, no. 9, Aug. 2021.


% System Model
\bibitem{arxiv2005} S. Kim and B. J. Kim, ``Novel backoff mechanism for mitigation of congestion in DSRC broadcast,'' \textit{arXiv preprint arXiv:2005.08921}, May 2020.

\bibitem{haengi05} M. Haenggi, ``On distances in uniformly random networks,'' \textit{IEEE Trans. Inf. Theory}, vol. 51, no. 10, Oct. 2005.

% Analysis
\bibitem{fire08} A. Singh, T. Das, P. Maniatis, P. Druschel, and T. Roscoe, ``BFT protocols under fire,'' in \textit{Proc. USENIX NSDI 2008}.

\bibitem{book} S. Chiu, D. Stoyan, W. Kendall, and J.Mecke, \textit{Stochastic Geometry and its Applications}, John Wiley \& Sons, Jun. 2013.

\bibitem{burke} O. Ibe, \textit{Markov Processes for Stochastic Modeling}, Newnes; May 2013.

\bibitem{gossip_access19} W. Hu, Y. Hu, W. Yao, and H. Li, ``A blockchain-based Byzantine consensus algorithm for information authentication of the internet of vehicles,'' \textit{IEEE Access}, Sep. 2019.

\bibitem{gossip_arxiv18} E. Buchman, J. Kwon, and Z. Milosevic, ``The latest gossip on BFT consensus,'' \textit{arXiv preprint arXiv:1807.04938}, Jul. 2018.

\bibitem{gossip11} M. Jelasity, \textit{Gossip}, Springer Berlin Heidelberg, pp. 139–162, [Online]. Available: \url{http://dx.doi.org/10.1007/978-3-642-17348-6_7}

\bibitem{scale18} C. Berger and H. P. Reiser, ``Scaling byzantine consensus: a broad analysis,'' in \textit{Proc. ACM SERIAL 2018}.

% Results
\bibitem{vtc21} S. Kim, B. J. Kim, and B. Park, ``Environment-adaptive multiple access for distributed V2X network: A reinforcement learning framework,'' \textit{arXiv preprint arXiv:2101.10447}, Jan. 2021.

\bibitem{access20} S. Kim and B. J. Kim, ``Crash risk-based prioritization of basic safety message in DSRC,'' \textit{IEEE Access}, vol. 8, Nov. 2020.

\end{thebibliography}
\end{document}